\documentclass[12pt]{iopart}
\usepackage{graphicx} 
\font\tenimbf=cmmib10 at 10pt
\font\sevenimbf=cmmib10 at 6pt
\font\fiveimbf=cmmib10 at 4pt
\newfam\imbf
\textfont\imbf=\tenimbf
\scriptfont\imbf=\sevenimbf
\scriptscriptfont\imbf=\fiveimbf

\def\bea{\begin{eqnarray}}
\def\eea{\end{eqnarray}}
\def\beq{\begin{equation}}
\def\eeq{\end{equation}}
\begin{document}

\title{Inclusive distributions at the LHC as predicted from the
    DPMJET-III model with chain fusion}

\author{F.Bopp$^1$, R.Engel$^2$  J.Ranft$^1$    and
S.Roesler$^3$}

\address{$^1$ Siegen University, Siegen, Germany}
\address{$^2$ Forschungszentrum Karlsruhe, Karlsruhe, Germany}
\address{$^3$ CERN, Geneva, Swtzerland}
\ead{Johannes.Ranft@cern.ch}

%
%
%


\begin{abstract}
    DPMJET-III with chain fusion is used to calculate inclusive
    distributions of Pb-Pb  collisions at LHC energies.
    We present  rapidity distributions as well as
    scaled multiplicities at mid-rapidity as function of the
    collision energy and the number of participants.
\end{abstract}

\section{Bibliography}

\maketitle 

Monte Carlo codes based on the two--component Dual Parton Model
(soft hadronic chains and hard hadronic collisions) are
available since 10--20 years:The present codes are
PHOJET for h--h and $\gamma$--h
collisions \cite{phojet-a} and DPMJET-III based on
PHOJET for h--A and A--A collisions \cite{dpmjet2}.
To apply
DPMJET--III to central collisions of heavy nuclei 
the percolation and fusion of the hadronic chains
had to be implemented \cite{dpmfusion1}.

\includegraphics[width=14cm]{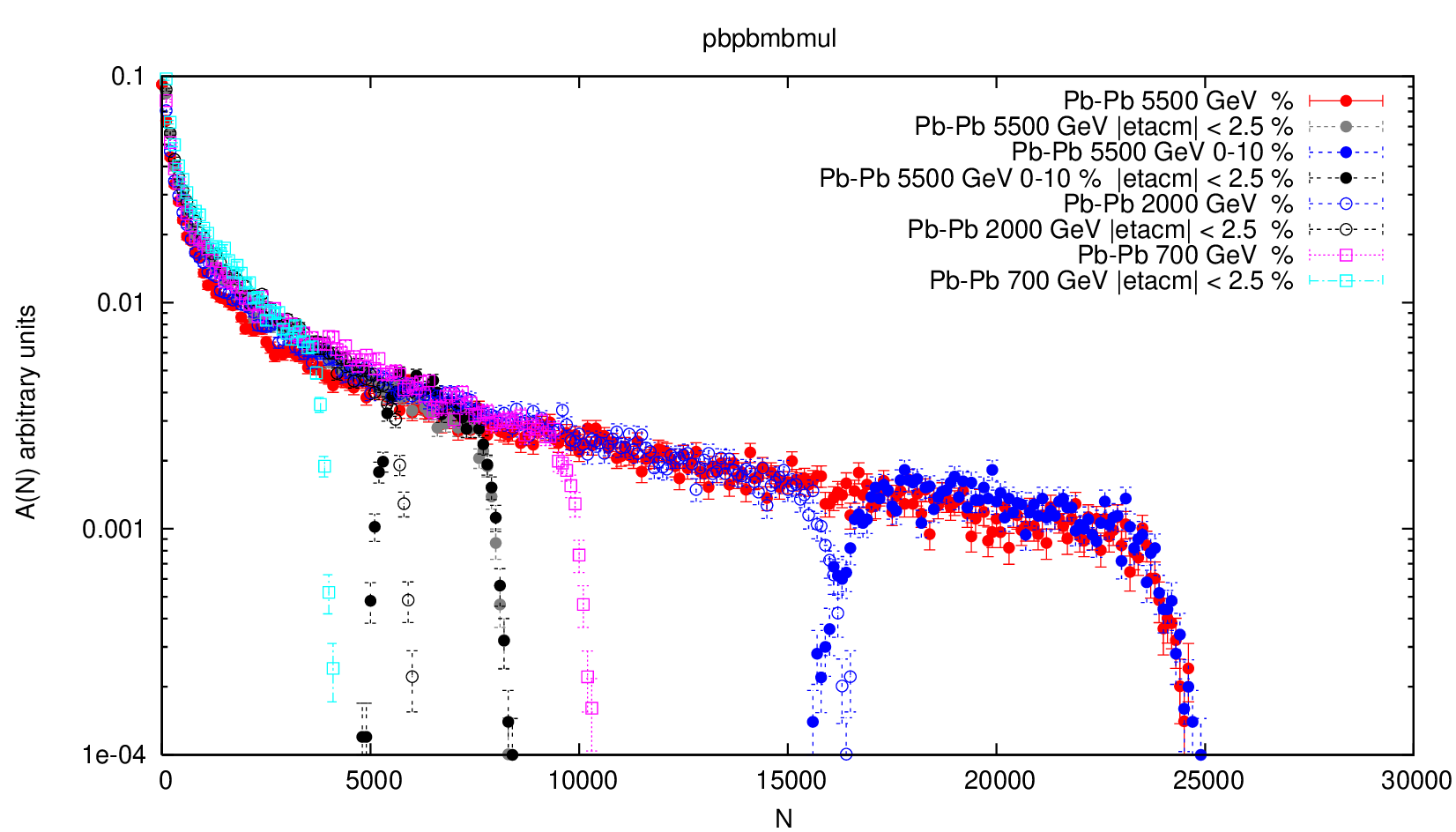}
\\
{{\bf Fig.1} Multiplicity distributions in minimum bias and
0--10 \% central collisions in Pb--Pb collisions in
the full $\eta_{cm}$ range and for  $|\eta_{cm}| \leq $  2.5.
(from DPMJET-III).}

\vskip 5mm

\includegraphics[width=7cm]{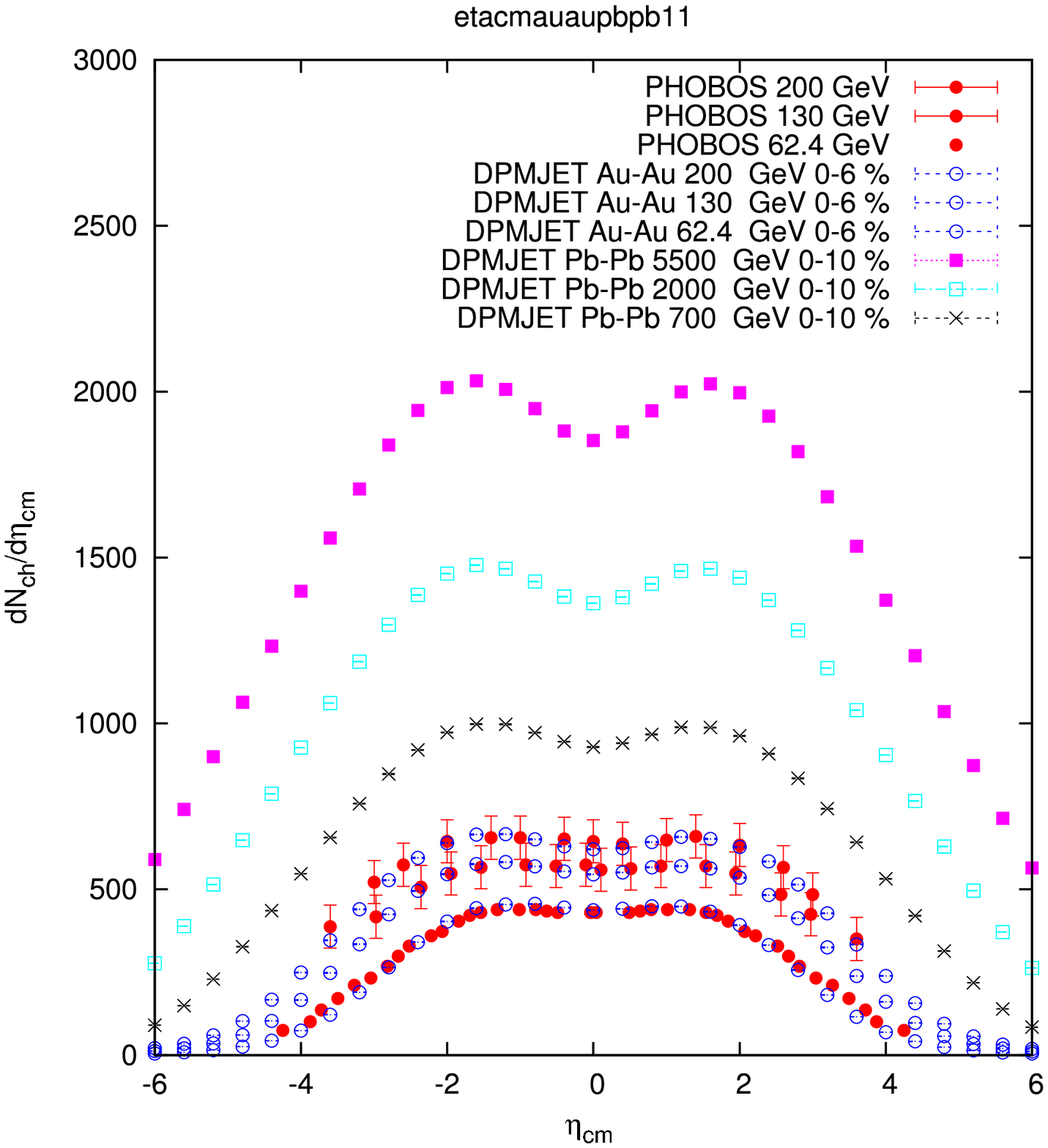}
\includegraphics[width=7cm]{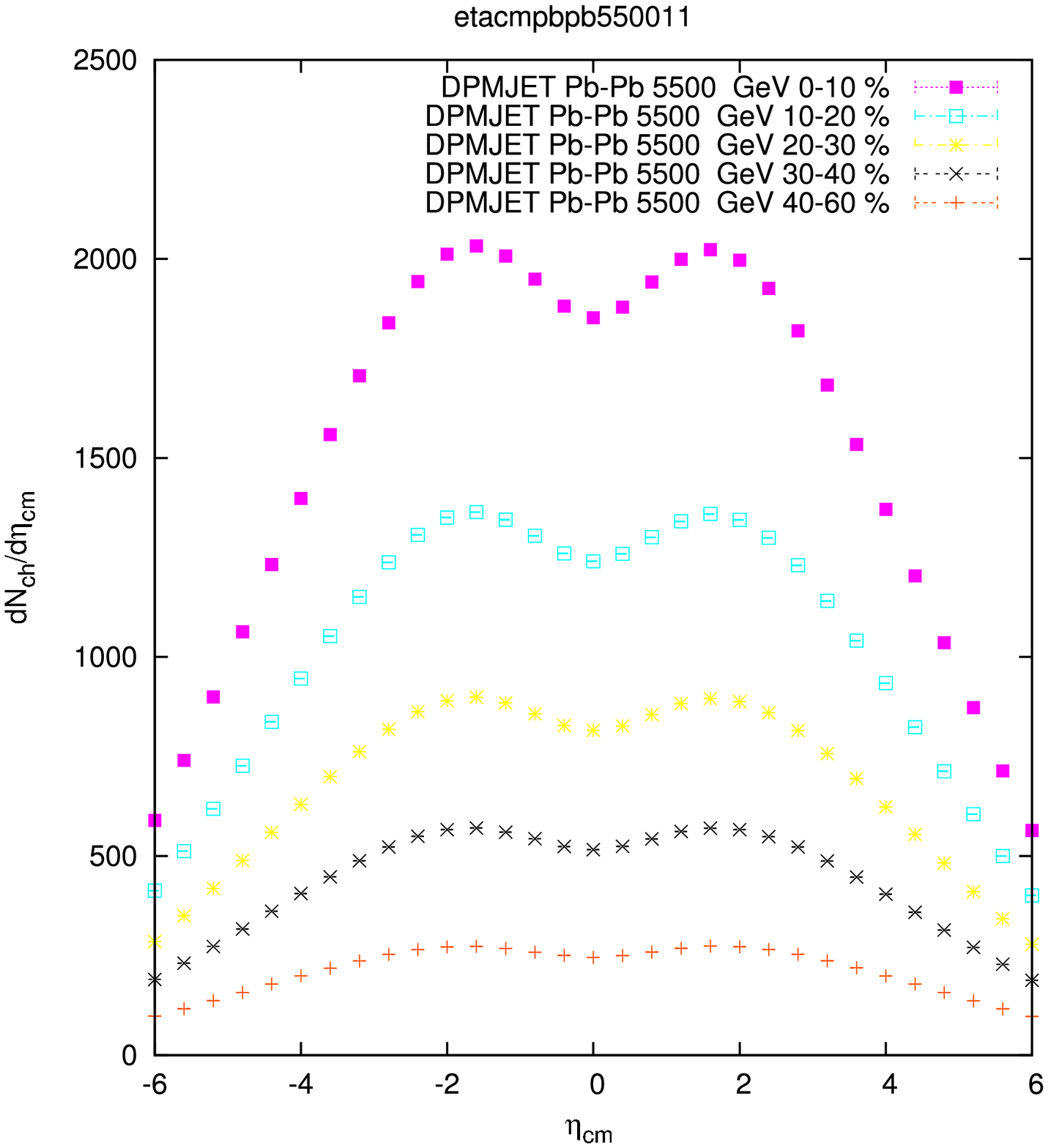}
\\
{{\bf Fig.2}(left) Central RHIC and LHC collisions (right) LHC Pb--Pb
collisions
from DPMJET-III.} 
\clearpage

In Figs.1 and 2 we apply this model
to minimum bias and central collisions of heavy nuclei at the LHC and 
at RHIC. We
find an excellent agreement to RHIC data on inclusive
distributions.

The behaviour of the inclusive hadron production becomes
particular simple if we plot it in the form
$\frac{dN}{d{\eta_{cm}}}/\frac{N_{part}}{2}$.  $N_{part}$ is the
number of participants in the A--A collisions. In Fig.3 we plot
this quantity as function of  $N_{part}$ and as function of
$E_{CM}$, in both plots we find a rather simple behaviour.

\includegraphics[width=7cm]{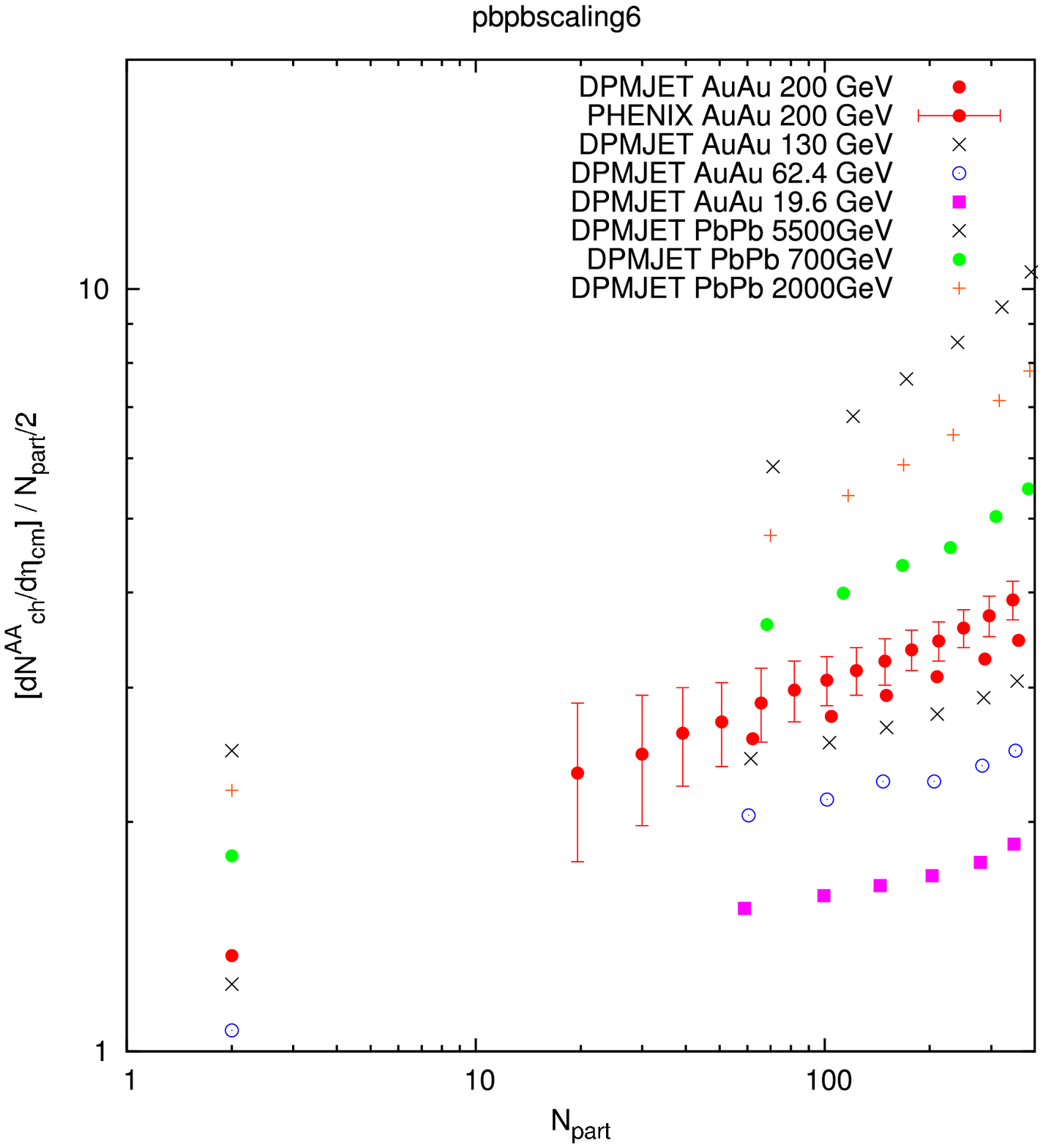}
\includegraphics[width=7cm]{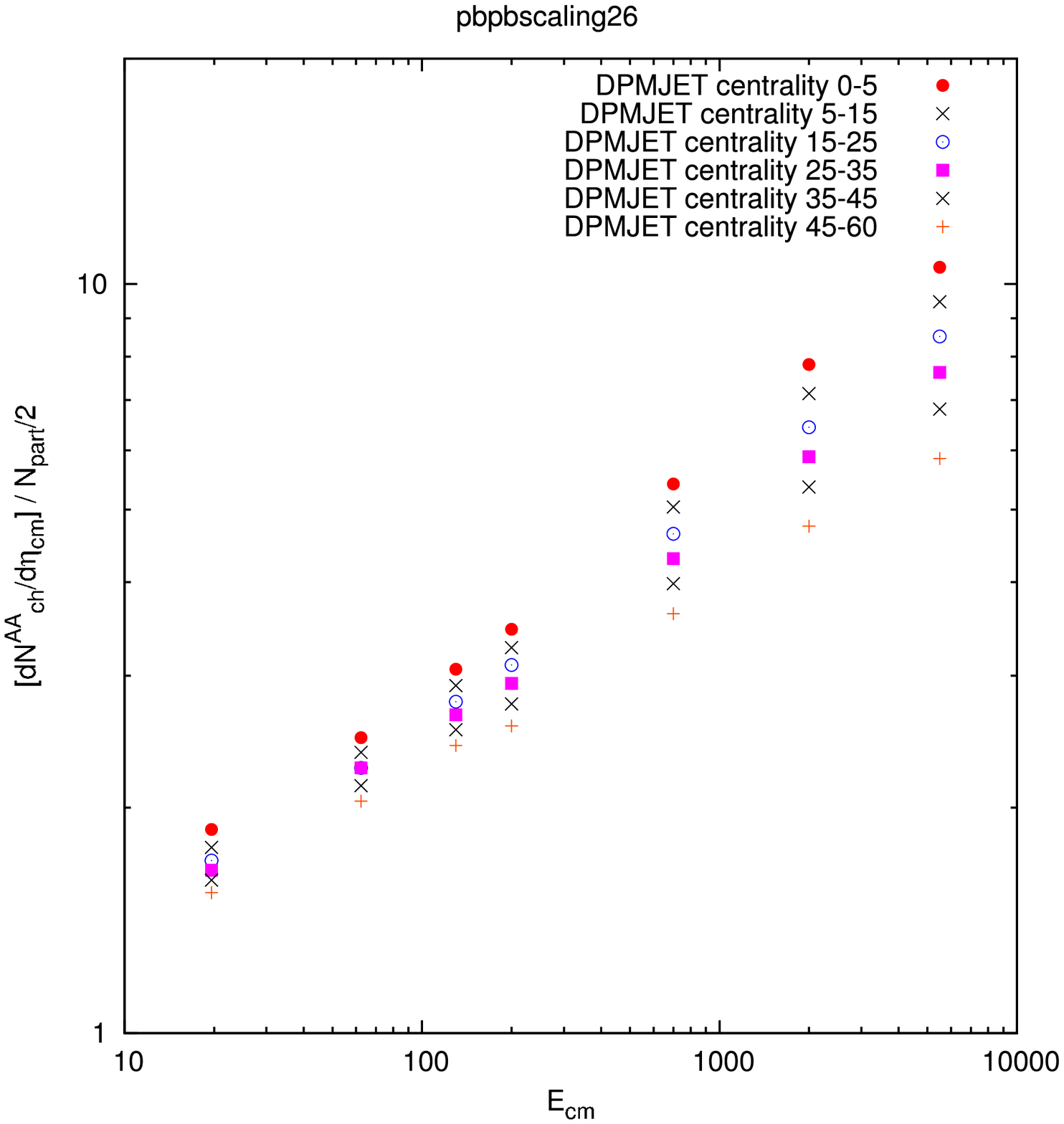}
\\
{{\bf Fig.3} $\frac{dN}{d{\eta_{cm}}}/\frac{N_{part}}{2}$
(left) over $N_{part}$ (right) over $E_{CM}$, Pb--Pb and Au--Au
collisions. 
} 

\vskip 3mm
The limiting fragmentation hypothesis was proposed in 1969 by
Benecke et al. \cite{limfrag}.
If we apply it to nuclear collisions we have to
plot $\frac{dN}{d{\eta_{cm}}}/\frac{N_{part}}{2}$ as function of 
 $\eta_{cm} - y_{beam}$. In Fig.4 we plot central and less
 central 
 Au--Au collisions at
 RHIC and LHC energies in this form. We find that DPMJET--III
 shows in the fragmentation region 
 only  small deviations from limiting fragmentation.

\vskip 3mm
\includegraphics[width=7cm]{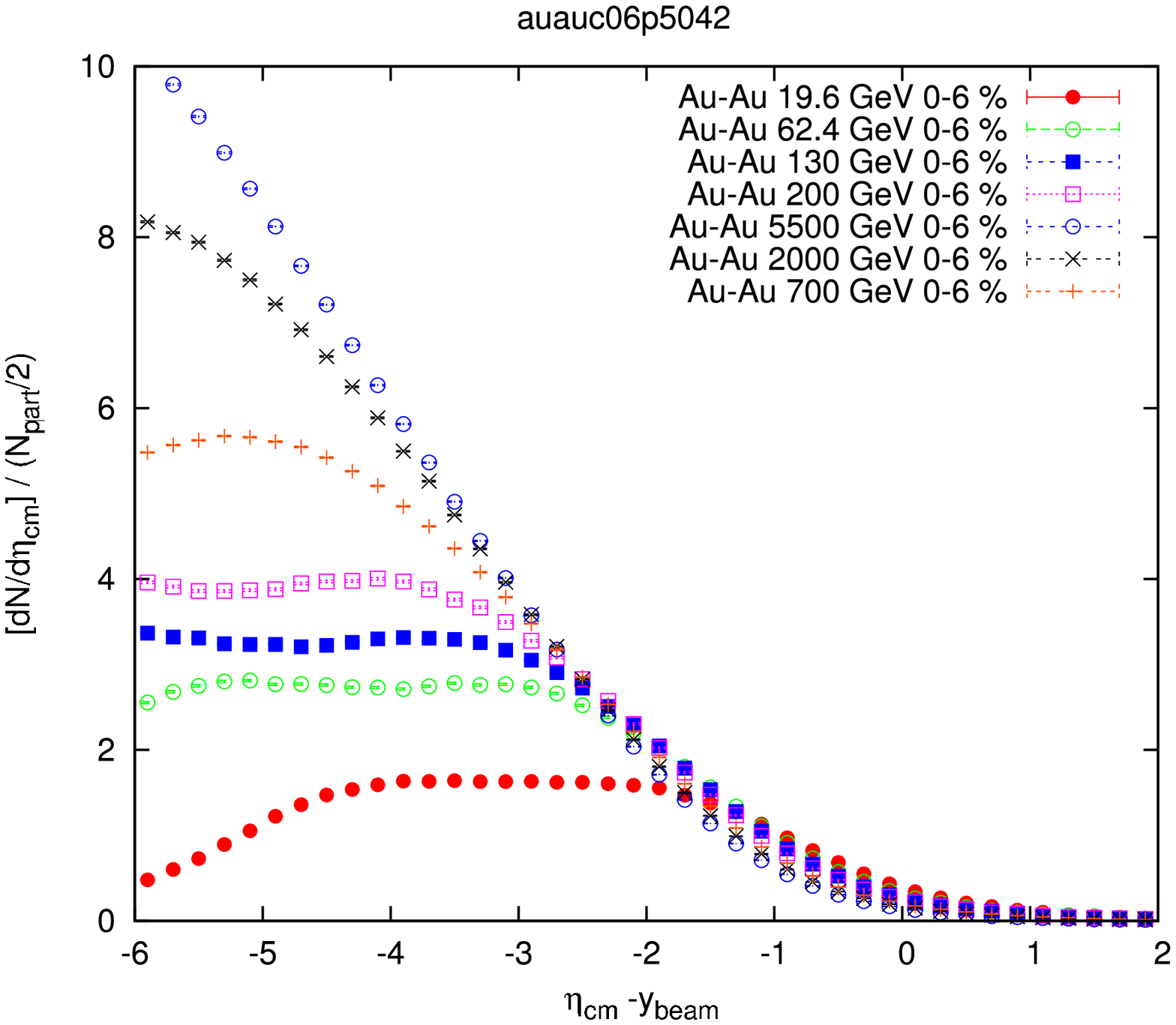}
\includegraphics[width=7cm]{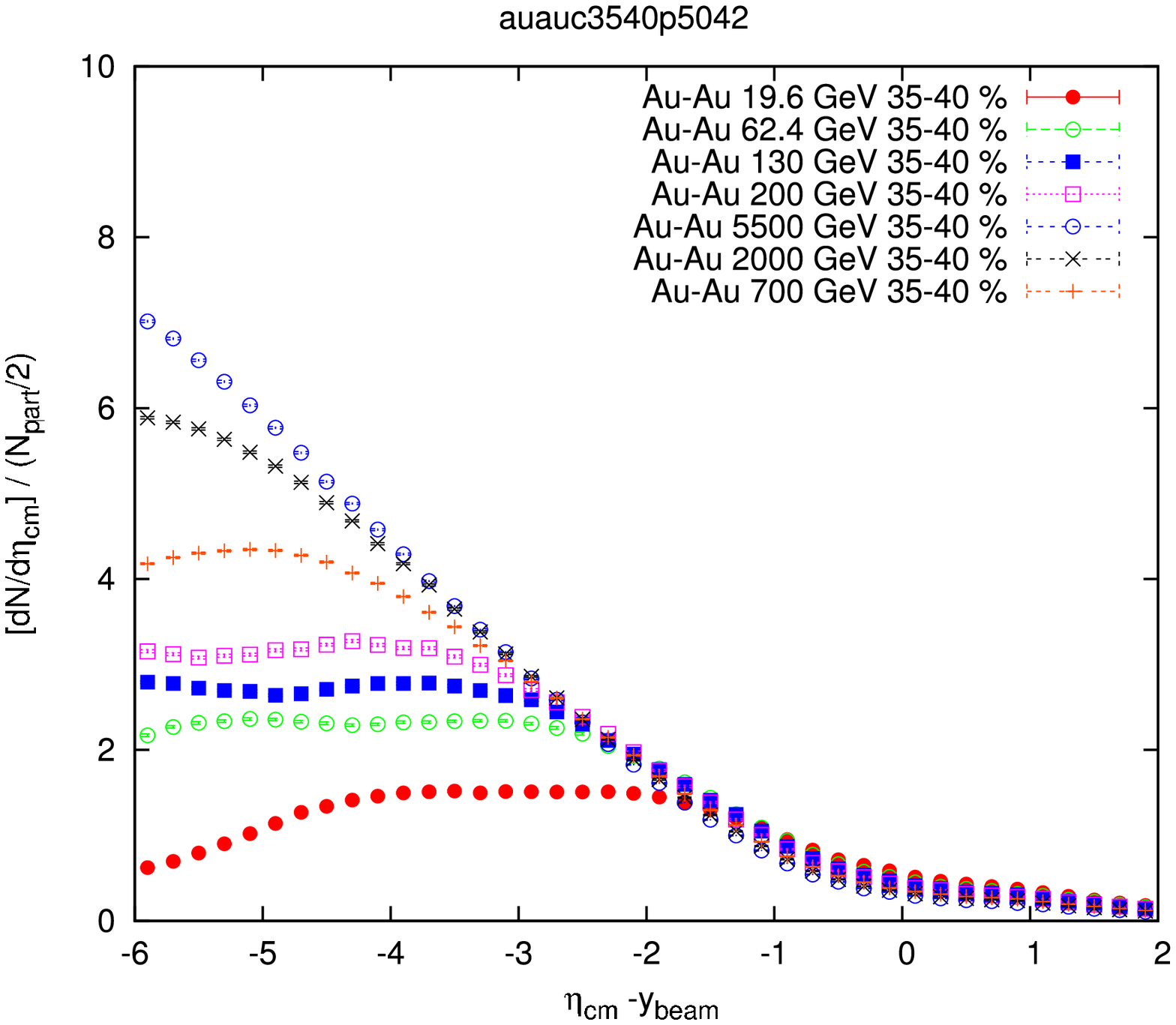}
\\
{{\bf Fig.4} $\frac{dN}{d{\eta_{cm}}}/\frac{N_{part}}{2}$ Au-Au
collisions over $\eta_{cm} - y_{beam}$
(left) central,  (right) less central. 
}

\end{document}